\documentclass[fleqn,11pt]{article}
\usepackage{cite}
\usepackage[pdftex]{hyperref}
\usepackage{amsmath, amsthm, amssymb, graphicx,slashed,mathrsfs}
\parskip=\baselineskip

\font\teneurm=eurm10 \font\seveneurm=eurm7 \font\fiveeurm=eurm5
\newfam\eurmfam
\textfont\eurmfam=\teneurm \scriptfont\eurmfam=\seveneurm
\scriptscriptfont\eurmfam=\fiveeurm

 \font\teneusm=eusm10 \font\seveneusm=eusm7 \font\fiveeusm=eusm5
\newfam\eusmfam
\textfont\eusmfam=\teneusm \scriptfont\eusmfam=\seveneusm
\scriptscriptfont\eusmfam=\fiveeusm

\font\tencmmib=cmmib10 \skewchar\tencmmib='177
\font\sevencmmib=cmmib7 \skewchar\sevencmmib='177
\font\fivecmmib=cmmib5 \skewchar\fivecmmib='177
\newfam\cmmibfam
\textfont\cmmibfam=\tencmmib \scriptfont\cmmibfam=\sevencmmib
\scriptscriptfont\cmmibfam=\fivecmmib

\parskip=\baselineskip

\def\p{\mbox{\boldmath$\displaystyle\mathbf{p}$}}

\def\bv{\mbox{\boldmath$\displaystyle\mathbf{\varphi}$}}
\def\hp{\mbox{\boldmath$\displaystyle\mathbf{\widehat{\p}}$}}
\def\0{\mbox{\boldmath$\displaystyle\mathbf{0}$}}
\def\s{\mbox{\boldmath$\displaystyle\mathbf{\sigma}$}}
\def\J{\mbox{\boldmath$\displaystyle\mathbf{J}$}}

\def\x{\mbox{\boldmath$\displaystyle\mathbf{x}$}}

\newcommand{\dual}[1]{\overset{{}^{{}^{\boldsymbol{\neg}}}}{\smash[t]{#1}}}

\begin{document}

\vskip 1.5in
\begin{center}
{\bf\Large{Elko in 1+1 dimensions}}\vskip 0.5cm
{Cheng-Yang Lee} \vskip 0.1in
{\small{
{\textit{Institute of Mathematics, Statistics and Scientific Computation,\\
Unicamp, 13083-859 Campinas, S\~{a}o Paulo, Brazil.\\
Email: cylee@ime.unicamp.br}}}}
\end{center}

\begin{abstract}
The quantum field operator for spin-half Elko describes a massive self-interacting fermionic dark matter candidate of mass dimension one. 
It has been shown that the theory has a built-in violation of the Lorentz symmetry and
a well-defined element of non-locality in the form of a preferred direction. This note shows that quantum field operators constructed using spin-half and higher-spin Elko violate Lorentz symmetry from first principle. Subsequently, we study the kinematics of Elko and its quantum field operator for any spin along the preferred direction.

\end{abstract}

\section{Introduction}  
The theory of Elko and spin-half fermionic field was an unexpected discovery made by Ahluwalia and Grumiller~\cite{Ahluwalia:2004ab,Ahluwalia:2004sz}. The field has mass dimension one instead of three-half so it allows renormalisable self-interaction. In addition, the field has other properties that makes it a dark matter candidate. 

Elko and its quantum field operator have attracted interests from many disciplines. In cosmology, various authors have shown that Elko is a potential inflaton candidate as well as a possible source of dark energy~\cite{Boehmer:2010ma,Boehmer:2010tv,Boehmer:2009aw,Boehmer:2008ah,Boehmer:2008rz,Boehmer:2007ut,
Boehmer:2007dh,Boehmer:2006qq,Chee:2010ju,Wei:2010ad,Shankaranarayanan:2010st,Shankaranarayanan:2009sz,
Gredat:2008qf,S.:2014dja,Pereira:2014wta,Basak:2014qea}. The gravitational interactions of Elko have been explored in various scenarios~\cite{Jardim:2014xla,daRocha:2014dla,Fabbri:2014foa}. The mathematical properties of Elko have been studied in detail by da Rocha et al.~\cite{HoffdaSilva:2009is,
daRocha:2008we,daRocha:2007pz,daRocha:2005ti,daRocha:2011yr,daRocha:2013qhu,Cavalcanti:2014uta,Bonora:2014dfa,Ablamowicz:2014rpa}. The signatures of mass dimension one fermions at the LHC as well as in the cosmological setting have also been investigated~\cite{Agarwal:2014oaa,Alves:2014kta}. In quantum field theory, Fabbri has shown that Elko does not violate causality ~\cite{Fabbri:2010va,Fabbri:2009aj,Fabbri:2009ka}. Wunderle and Dick have used Elko to construct
supersymmetric Lagrangians for fermionic fields with mass dimension one~\cite{Wunderle:2010yw}.

In refs.~\cite{Ahluwalia:2008xi,Ahluwalia:2009rh}, it was shown that the fermionic field is only local along the 3-axis. Therefore, the theory has a preferred direction and thus violates Lorentz symmetry. A possible solution was suggested by Ahluwalia and Horvath~\cite{Ahluwalia:2010zn}. Their results suggest that the theory is invariant under very special relativity proposed by Cohen and Glashow~\cite{Cohen:2006ky}. For a comprehensive review of the theory, please refer to ref.~\cite{Ahluwalia:2013uxa}.

In this note, following the formalism developed by Wigner~\cite{Wigner:1939cj} and Weinberg~\cite{Weinberg:1964cn,Weinberg:1964ev,Weinberg:1995mt}, we show from first principle that quantum field operators constructed from Elko of any spin violate Lorentz symmetry. 
Subsequently, we study the kinematics of Elko and its quantum field operators for any spin along the preferred direction. 

\section{The Elko consruct}

In this section, we first review the spin-half Elko construct, its higher-spin generalisation and the existence of the preferred direction. We then prove that the corresponding quantum field operators violate Lorentz symmetry.

In the $(\frac{1}{2},0)\oplus(0,\frac{1}{2})$ representation space, Elko is constructed from the following four-component spinor of the form
\begin{equation}
\chi(\p,\sigma)=\sqrt{m}\left[\begin{array}{cc}
\vartheta\,\Theta\phi^{*}(\p,\sigma) \\
\phi(\p,\sigma)\end{array}\right] \label{eq:elko_spinors}
\end{equation}
where $\vartheta$ is a phase to be determined and $\Theta$ is the spin-half Wigner time-reversal operator,
\begin{equation}
\Theta=\left(\begin{array}{cc}
0 & -1 \\
1 & 0 \end{array}\right).\label{eq:time}
\end{equation}
Generally, it is defined as
\begin{equation}
\J=-\Theta\J^{*}\Theta
\end{equation}
where $\J$ is the rotation generator of the $(j,0)$ and $(0,j)$ representations. The use of the Wigner time-reversal operator allows us to generalise Elko to higher-spin.

In eq.~(\ref{eq:elko_spinors}), $\phi(\p,\sigma)$ is a left-handed Weyl spinor. The boost transformation is given by
\begin{equation}
\phi(\p,\sigma)=\exp\left(-\frac{\s}{2}\cdot\bv\right)\phi(\boldsymbol{\epsilon},\sigma)\label{eq:spinor_boost}
\end{equation}
where the rapidity parameter $\bv=\varphi\hp$ is defined as 
\begin{equation}
\cosh{\varphi}=\frac{E_{\mathbf{p}}}{m},\hspace{0.5cm} \sinh{\varphi}=\frac{|\p|}{m}.
\end{equation}
The rest spinor $\phi(\boldsymbol{\epsilon},\sigma)$ is chosen to be eigenspinors of the helicity operator
\begin{equation}
\frac{1}{2}\s\cdot\hat{\p}\,\phi(\boldsymbol{\epsilon},\sigma)=\sigma
\phi(\boldsymbol{\epsilon},\sigma)
\end{equation}
with $\boldsymbol{\epsilon}=\lim_{\mathbf{p}\rightarrow\mathbf{0}}\hat{\p}$. From eq.~(\ref{eq:spinor_boost}), a straightforward calculation shows that $\vartheta\Theta\phi^{*}(\p,\sigma)$  transforms as a right-handed Weyl spinor~\cite{Ahluwalia:2004sz} 
\begin{equation}
\vartheta\Theta\phi^{*}(\p,\sigma)=\exp\left(\frac{\s}{2}\cdot\bv\right)[\vartheta\Theta\phi^{*}(\boldsymbol{\epsilon},\sigma)].
\end{equation}
The spinor of arbitrary momentum is then related to its rest spinor by
\begin{equation}
\chi(\p,\sigma)=\kappa(\p)\chi(\boldsymbol{\epsilon},\sigma)
\end{equation}
where
\begin{equation}
\kappa(\p)=\left[\begin{array}{cc}
\exp\left(\frac{\s}{2}\cdot\bv\right) & \mbox{O} \\
\mbox{O} &\exp\left(-\frac{\s}{2}\cdot\bv\right)\end{array}\right].\label{eq:boost}
\end{equation}

The phase $\vartheta$ is determined so that $\chi(\p,\sigma)$ is an eigenspinor of charge-conjugation operator $\mathcal{C}$,
\begin{equation}
\mathcal{C}=\left(\begin{array}{cc}
\mbox{O} & -i\Theta^{-1} \\
-i\Theta & \mbox{O} \end{array}\right)K
\end{equation}
where $K$ complex conjugates everything on its right. Therefore, we get
\begin{equation}
\mathcal{C}\chi(\p,\sigma)\vert_{\vartheta=\pm i}=\pm \chi(\p,\sigma)\vert_{\vartheta=\pm i}. \label{eq:c_chi}
\end{equation}
Equation~(\ref{eq:c_chi}) gives us four Elkos. They are defined as
\begin{equation}
\mathcal{C}\xi(\p,\sigma)=\xi(\p,\sigma),\hspace{0.5cm}
\mathcal{C}\zeta(\p,\sigma)=-\zeta(\p,\sigma).
\end{equation}
The spinors with positive and negative eigenvalues are referred to as the self-conjugate and anti-self-conjugate spinors respectively. 

The above construction of spin-half Elko can be generalised to higher-spin using the fact that up to a constant, the Wigner time-reversal operator $\Theta^{(j)}$ for the $(j,0)$ and $(0,j)$ representations is given by~\cite{Lee:2012td}
\begin{equation}
\Theta^{(j)}_{\bar{\sigma}\sigma}=(-1)^{-j-\bar{\sigma}}\delta_{-\bar{\sigma},\sigma}.\label{eq:Theta_soln}
\end{equation}
The constant is chosen such that when $j=\frac{1}{2}$, it yields eq.~(\ref{eq:time}). The higher-spin Elko takes the form
\begin{equation}
\chi^{(j)}(\p,\sigma)=\left[
\begin{matrix}
\vartheta\Theta^{(j)}\phi^{(j)*}(\p,\sigma)\\
\phi^{(j)}(\p,\sigma)
\end{matrix}\right]\label{eq:elko_any_spin}
\end{equation}
where $\phi^{(j)}(\p,\sigma)$ is a $(2j+1)$-component eigenfunction of $\mathbf{J\cdot\hat{p}}$ with eigenvalues $\sigma=-j,\cdots j$.
  One interesting result to note is that when $j=1,2,\cdots$, the self-conjugate and anti-self conjugate spinors have negative and positive eigenvalues respectively. For more details on their construction, please see ref.~\cite{Lee:2012td}.

The quantum field operator $\Lambda(x)$ and its adjoint $\dual{\Lambda}(x)$ with Elko and its dual of any spin as expansion coefficients are defined as\footnote{We use the notation $p\cdot x=p^{\mu}x_{\mu}$ where the metric is defined as $\eta^{\mu\nu}=\mbox{diag}(1,-1,-1,-1)$.}
\begin{subequations}
\begin{equation}
\Lambda(x)=(2\pi)^{-3/2}\int\frac{d^{3}p}{\sqrt{2mE_{\mathbf{p}}}}\sum_{\sigma}\big[e^{-ip\cdot x}\xi(\p,\sigma)a(\p,\sigma)+e^{ip\cdot x}\zeta(\p,\sigma)b^{\dag}(\p,\sigma)\big],
\end{equation}
\begin{equation}
\dual{\Lambda}(x)=(2\pi)^{-3/2}\int\frac{d^{3}p}{\sqrt{2mE_{\mathbf{p}}}}\sum_{\sigma}\big[e^{ip\cdot x}\dual{\xi}(\p,\sigma)a^{\dag}(\p,\sigma)+e^{-ip\cdot x}\dual{\zeta}(\p,\sigma)b(\p,\sigma)\big]
\end{equation}
\end{subequations}
where $a(\p,\sigma)$ and $b(\p,\sigma)$ are the annihilation operators for the particles and anti-particles respectively. They satisfy the standard commutator $(j=1=2,\cdots)$ and anti-commutator $(j=\frac{1}{2},\frac{3}{2},\cdots)$ algebras
\begin{eqnarray}
[a(\p,\sigma),a^{\dag}(\p',\sigma')]_{\pm}&=&[b(\p,\sigma),b^{\dag}(\p',\sigma')]_{\pm}\label{eq:anticommutation} \\
&=&\delta_{\sigma\sigma'}\delta^{3}(\p-\p').\nonumber
\end{eqnarray}
The solutions for spin-half Elko, its dual and higher-spin generalisation can be found in ref.~\cite{Ahluwalia:2013uxa} and ref.~\cite{Lee:2012td} respectively. Analogous to the Majorana fermions~\cite{Majorana:1937vz}, we can also construct another field operator $\lambda(x)$,
\begin{equation}
\lambda(x)=\Lambda(x)\big\vert_{b^{\dag}=a^{\dag}}
\end{equation}
where the particles are identical to the antiparticles. 

The element of non-locality is encoded the commutator/anti-commutators $[\Lambda(t,\mathbf{x}),\Pi(t,\mathbf{y})]_{\pm}$ and $[\Lambda(t,\mathbf{x}),\dual{\Lambda}(t,\mathbf{y})]_{\pm}$  where
\begin{equation}
\Pi(x)=\frac{\partial\dual{\Lambda}}{\partial t}(x)
\end{equation}
is obtained from the Klein-Gordon Lagrangian density. A straightforward calculation shows that~\cite{Lee:2012td}
\begin{subequations}
\begin{equation}
[\Lambda(t,\mathbf{x}),\Pi(t,\mathbf{y})]_{+}=i\int\frac{d^{3}p}{(2\pi)^{3}}e^{i\mathbf{p\cdot(x-y)}}[I+\mathcal{G}(\phi)]\nonumber
\end{equation}
\begin{equation}
[\Lambda(t,\mathbf{x}),\dual{\Lambda}(t,\mathbf{y})]_{+}=O,\hspace{0.5cm}j=\frac{1}{2},\frac{3}{2},\cdots\label{eq:anti},
\end{equation}
\begin{equation}
[\Lambda(t,\mathbf{x}),\Pi(t,\mathbf{y})]_{-}=i\int\frac{d^{3}p}{(2\pi)^{3}}e^{i\mathbf{p\cdot(x-y)}}
\mathcal{G}(\phi)\nonumber
\end{equation}
\begin{eqnarray}
[\Lambda(t,\x),\dual{\Lambda}(t,\mathbf{y})]_{-}&=&(2\pi)^{-3}\int\frac{d^{3}p}{4E_{\mathbf{p}}}e^{i\mathbf{p\cdot(x-y)}}I \nonumber\\
&=&\frac{m}{8\pi^{2}|\x-\mathbf{y}|}K_{1}(m|\x-\mathbf{y}|),\hspace{0.5cm}j=1,2,\cdots.\label{eq:com}
\end{eqnarray}
\end{subequations}
where $K_{1}(m|\x|)$ is the standard Hankel function. The general solution of $\mathcal{G}(\phi)$ is given by
\begin{equation}
\mathcal{G}(\phi)=\beta(j)(-1)^{2j}
\left[\begin{matrix}
\mbox{O} & G(\phi) \\
G(\phi) & \mbox{O}
\end{matrix}\right] \label{eq:G}
\end{equation}
where 
\begin{equation}
\beta(j)=
\begin{cases}
-i & j=\frac{1}{2},\frac{3}{2},\cdots \\
1 & j=1,2,\cdots
\end{cases}
\end{equation}
and the sub-matrix $G(\phi)$ is
\begin{equation}
G_{\ell m}(\phi)=(-1)^{j+\ell}e^{-2i\ell\phi}\delta_{\ell,-m}.
\end{equation}
In eqs.~(\ref{eq:anti}) and (\ref{eq:com}), the integrals over $\mathcal{G}(\phi)$ only vanishes when $\mathbf{x-y}$ is aligned to the 3-axis thus indicating the existence of a preferred direction and Lorentz violation. Note that along the 3-axis, the fermionic fields satisy the canonical algebras but the bosonic fields remain non-local.


We note, although Elko transforms correctly under the $(j,0)\oplus(0,j)$ representation, this does not guarantee that the corresponding quantum field operators $\Lambda(x)$ and $\lambda(x)$ satisfy Lorentz symmetry. Indeed, the existence of a preferred direction already indicate that the symmetry is violated. The works of Weinberg~\cite{Weinberg:1964cn,Weinberg:1964ev,Weinberg:1995mt} provide a more rigorous proof of Lorentz violation. Given a representation of the Lorentz group, the demand of Lorentz symmetry put constraints on the expansion coefficients of the quantum field operator up to a few caveats (Wigner type degeneracy~\cite{Wigner:1939cj}). The field equation can then be derived from the properties of the associated expansion coefficients and the propagator. 

If the quantum field operators $\Lambda(x)$ and $\lambda(x)$ are Lorentz-invariant, then the expansions coefficients $\xi(\p,\sigma)$ and $\zeta(\p,\sigma)$ should satisfy the constraints imposed by Lorentz symmetry. However, this is not the case. We now prove this for Elko and its higher-spin generalisation.  Recalling that $\J$ is the rotation generator of the $(j,0)$ and $(0,j)$ representations, rotation symmetry about the 3-axis requires the coefficients at rest to satisfy the following equations~\cite[sec.~5.1]{Weinberg:1995mt}
\begin{subequations}
\begin{equation}
\sum_{\bar{\sigma}}\xi_{\ell}(\0,\bar{\sigma})(J_{3})_{\bar{\sigma}\sigma}=\sum_{\bar{\ell}}(\mathcal{J}_{3})_{\ell\bar{\ell}}\,\xi_{\bar{\ell}}(\0,\sigma), \label{eq:rot1}
\end{equation}
\begin{equation}
\sum_{\bar{\sigma}}\zeta_{\ell}(\0,\bar{\sigma})(J_{3})^{*}_{\bar{\sigma}\sigma}=-\sum_{\bar{\ell}}(\mathcal{J}_{3})_{\ell\bar{\ell}}\,\zeta_{\bar{\ell}}(\0,\sigma) \label{eq:rot2}
\end{equation}
\end{subequations}
where $\ell=1,\cdots,2\times(2j+1)$ denotes the components of the coefficients and $\mathcal{J}_{3}$ is the rotation generator of the $(j,0)\oplus(0,j)$ representation about the 3-axis
\begin{equation}
\mathcal{J}_{3}=
\left(\begin{array}{cc}
J_{3} & O  \\
O & J_{3} \end{array}\right). \label{eq:rot_generator}
 \end{equation}
The Elkos for any spin take the form of eq.~(\ref{eq:elko_any_spin}). An important point to note is that $\xi(\0,\sigma)$ and $\zeta(\0,\sigma)$ are actually defined in the polarisation basis with $\phi^{(j)}(\0,\sigma)$ being an eigenfunction of $J_{3}$ instead of $\mathbf{J\cdot\hat{p}}$. But since $\phi^{(j)}(\boldsymbol{\epsilon},\sigma)$ and $\phi^{(j)}(\0,\sigma)$ are related by a unitary transformation, we may work in the polarisation basis where the calculations are easier to perform. Substitute Elkos into eqs.~(\ref{eq:rot1}) and (\ref{eq:rot2}), we find that these conditions are not satisfied. Therefore, the quantum field operators with Elko of any spin as expansion coefficients violate Lorentz symmetry.\footnote{This result was also obtained by Gillard and Martin in~\cite{Gillard:2010nr}.}


\section{Elko along the preferred direction}
We now turn our attention to investigate the theory along the preferred direction. The reasons for doing so will be explained shortly. In 3+1 dimensions, the solutions of spin-half Elko are given in the helicity basis~\cite{Ahluwalia:2013uxa}. Along the preferred direction, these solutions and their higher-spin generalisations, are obtained by taking the limit $\theta\rightarrow0$, $\phi\rightarrow0$ where $\theta$ and $\phi$ are the polar and azimuthal angles respectively. For example, the solutions for the spin-half Elkos are
\begin{equation}
\xi(0,\textstyle{\frac{1}{2}})=\sqrt{m}\left(\begin{array}{cccc}
0 \\
i \\
1 \\
0 \\ \end{array}\right),\hspace{0.3cm}
\xi(0,-\textstyle{\frac{1}{2}})=\sqrt{m}\left(\begin{array}{cccc}
-i \\
0 \\
0 \\
1 \\ \end{array}\right),\nonumber
\end{equation}
\begin{equation}
\zeta(0,\textstyle{\frac{1}{2}})=\sqrt{m}\left(\begin{array}{cccc}
i \\
0 \\
0 \\
1 \\ \end{array}\right),\hspace{0.3cm}
\zeta(0,-\textstyle{\frac{1}{2}})=\sqrt{m}\left(\begin{array}{cccc}
0 \\
i \\
-1 \\
0 \\ \end{array}\right). \label{eq:elko_like_spinors}
\end{equation}
The corresponding quantum field operator and its adjoint for any spin are given by
\begin{subequations}
\begin{equation}
\Lambda(t,x)=(2\pi)^{-1/2}\int\frac{dp}{\sqrt{2mE_{p}}}\sum_{\sigma}\big[e^{-ip\cdot x}\xi(p,\sigma)a(p,\sigma)+e^{ip\cdot x}\zeta(p,\sigma)b^{\dag}(p,\sigma)\big], \label{eq:field}
\end{equation}
\begin{equation}
\dual{\Lambda}(t,x)=(2\pi)^{-1/2}\int\frac{dp}{\sqrt{2mE_{p}}}\sum_{\sigma}\big[e^{ip\cdot x}
\dual{\xi}(p,\sigma)a^{\dag}(p,\sigma)+e^{-ip\cdot x}\dual{\zeta}(p,\sigma)b(p,\sigma)\big]
\label{eq:adjoint}
\end{equation}
\end{subequations}
where the expansion coefficients of finite momentum are obtained by boosting along the preferred direction (the 3-axis). The spin-sums defined with respect to the Elko duals are given by
\begin{subequations}
\begin{equation}
N(p)=\sum_{\sigma}\xi(p,\sigma)\dual{\xi}(p,\sigma)=m(\mathcal{G}+I), \label{eq:spin_sum1}
\end{equation}
\begin{equation}
M(p)=\sum_{\sigma}\zeta(p,\sigma)\dual{\zeta}(p,\sigma)=m(\mathcal{G}-I) \label{eq:spin_sum2}
\end{equation}
\end{subequations}
where $\mathcal{G}$ is a constant off-diagonal matrix obtained by taking the limit $\phi\rightarrow0$ of eq.~(\ref{eq:G})
\begin{equation}
\mathcal{G}=\lim_{\phi\rightarrow0}\mathcal{G}(\phi).
\end{equation}

There are two main reasons why we want to study the theory in 1+1 dimensions along the preferred direction. The first reason is to better understand the locality structure and the effects of Lorentz violation. The second reason is due to the observation that the Elko spin-sums given in eqs.~(\ref{eq:spin_sum1}) and (\ref{eq:spin_sum2}) are momentum-independent. This is different to their counterparts in 3+1 dimensions as well as the properties of the Dirac spinors, where the spin-sums contain angular and explicit momentum dependence respectively. We shall now explore the consequences as a result of confining the theory to the preferred direction.

First, we compute the propagator of the theory. The propagator is obtained from the time-ordered product of the vacuum expectation value
\begin{eqnarray}
S(x^{\mu},y^{\mu})&=&\langle\,\,|T[\Lambda(t,x)\dual{\Lambda}(t',y)]|\,\,\rangle \nonumber\\
&=&\theta(t-t')\langle\,\,|\Lambda(t,x)\dual{\Lambda}(t',y)|\,\,\rangle\pm
\theta(t'-t)\langle\,\,|\dual{\Lambda}(t',y)\Lambda(t,x)|\,\,\rangle
\end{eqnarray}
where $\theta(t)$ is the step function and the top and bottom signs are for the bosonic and fermionic fields respectively. Expand the propagator using eqs.~(\ref{eq:field}) and (\ref{eq:adjoint}), we obtain
\begin{subequations}
\begin{equation}
S(x^{\mu},y^{\mu})=i\int\frac{d^{2}p}{(2\pi)^{2}}e^{-ip\cdot(x-y)}\left[\frac{I+(p^{0}/E_{p})\,\mathcal{G}}{p\cdot p-m^{2}+i\epsilon}\right],\hspace{0.5cm}j=\frac{1}{2},\frac{3}{2},\cdots
\end{equation}
\begin{equation}
S(x^{\mu},y^{\mu})=i\int\frac{d^{2}p}{(2\pi)^{2}}e^{-ip\cdot(x-y)}\left[\frac{(p^{0}/E_{p})\,I+\mathcal{G}}{p\cdot p-m^{2}+i\epsilon}\right],\hspace{0.5cm}j=1,2,\cdots
\end{equation}
\end{subequations}
where $p^{0}$ is the off-shell time-component of the momentum. 
By performing a contour integration on the terms proportional to $p^{0}/E_{p}$, we find that the propagator for any spin can be written as
\begin{eqnarray}
S(x^{\mu},y^{\mu})&=&\pm i\int\frac{d^{2}p}{(2\pi)^{2}}e^{-ip\cdot(x-y)}\left[\frac{I+\mathcal{G}}{p\cdot p-m^{2}+i\epsilon}\right]\nonumber\\
&=&\pm(I+\mathcal{G})S_{\mbox{\tiny{KG}}}(x^{\mu},y^{\mu}),
\end{eqnarray}
where the top and bottom signs apply when $t-t'>0$ and $t'-t<0$ respectively and $S_{\mbox{\tiny{KG}}}(x^{\mu},y^{\mu})$ is the Klein-Gordon propagator of a massive scalar field. Therefore, up to a constant matrix, the propagator along the preferred direction is a Green's function of the Klein-Gordon operator. We may then take the free Lagrangian density for the theory to be
\begin{equation}
\mathscr{L}_{\Lambda}=\partial^{\mu}\dual{\Lambda}\partial_{\mu}\Lambda-m^{2}\dual{\Lambda}\Lambda.
\label{eq:Lagrangian}
\end{equation}
This time, because the spin-sums are momentum-independent, a straightforward computation shows that the theory satisfies the canonical algebra (up to a constant matrix)
\begin{subequations}
\begin{eqnarray}
[\Lambda(t,x),\Pi(t,y)]_{+}&=&i\int\frac{dp}{4\pi m}e^{ip(x-y)}[N(p)-M(-p)]\nonumber\\
&=&i\delta(x-y)I,\hspace{0.5cm}j=\frac{1}{2},\frac{3}{2},\cdots,
\end{eqnarray}
\begin{equation}
[\Lambda(t,x),\Pi(t,y)]_{-}=i\delta(x-y)\mathcal{G},\hspace{0.5cm}j=1,2,\cdots.
\end{equation}
\end{subequations}
But on the other hand, for the same reason, the following anti-commutators are non-vanishing at space-like separation
\begin{subequations}
\begin{eqnarray}
[\Lambda(t,x),\dual{\Lambda}(t,y)]_{+}&=&\int\frac{dp}{4\pi mE_{p}}e^{ip(x-y)}[N(p)+M(-p)]\nonumber\\
&=&\int\frac{dp}{2\pi E_{p}}e^{ip(x-y)}\mathcal{G} \nonumber\\
&=&\frac{1}{\pi}K_{0}(m|x-y|)\mathcal{G},\hspace{0.5cm} j=\frac{1}{2},\frac{3}{2},\cdots \label{eq:non_local1}
\end{eqnarray}
\begin{equation}
[\Lambda(t,x),\dual{\Lambda}(t,y)]_{-}=\frac{1}{\pi}K_{0}(m|x-y|)I,\hspace{0.5cm} j=1,2,\cdots.
\label{eq:non_local2}
\end{equation}
\end{subequations}
where $K_{0}(x)$ is the modified Bessel function of the second kind. This anti-commutator only vanishes in the limit $|x-y|\rightarrow\infty$.

In summary, this note shows that quantum field operators constructed with Elko of any spin violate Lorentz symmetry. The Lagrangian density and the locality structure along the preferred direction are different from their counterparts in 3+1 dimensions. For the later, it was suggested in ref.~\cite{Lee:2014opa} that the Klein-Gordon Lagrangian density must be modified in order to provide an adequate description of the spin-half fermions whereas the above analysis shows that no modifications are necessary. However, the non-localities exhibited in eqs.~(\ref{eq:non_local1}) and (\ref{eq:non_local2}) are problematic. If one wishes for the theory, or more precisely the $S$-matrix to preserve causality with interactions as functions of $\dual{\Lambda}(t,x)\Lambda(t,x)$, then eqs.~(\ref{eq:non_local1}) and (\ref{eq:non_local2}) must identically vanish. Note that this is also an issue for the bosonic but not the fermionic fields in 3+1 dimensions. Our results suggest that there are no causality-preserving interacting quantum field theories associated with Elko along the preferred direction. Nevertheless, there remains many avenues to explore in the classical realms. For  instance, since the field is described by a Klein-Gordon Lagrangian density and is therefore dimensionless, one could extend the theory in analogy to the non-linear $\sigma$-model. Just like the scalar field theories, we expect such extensions to have many interesting properties.


\section*{Acknowledgement}
This research is supported by CNPq grant 313285/2013-6. I am grateful to the hospitality of the Department of Physics and Astronomy at the University of Canterbury where part of this work was completed.

\label{Bibliography}
\bibliographystyle{JHEP}  
\bibliography{Bibliography}  
\end{document}